%% file: paper_v2.tex
\documentclass[conference]{IEEEtran}
\usepackage{color}
\usepackage{flushend}

\ifCLASSINFOpdf
  \usepackage[pdftex]{graphicx}
  \graphicspath{{scripts/}}
\else
  \usepackage[dvipsone]{graphicx}
  \graphicspath{{scripts/}}
  \DeclareGraphicsExtensions{.ps}
\fi
\usepackage{amsmath}
\usepackage{subfigure}

\begin{document}
%
\title{An Accurate Gaussian Process-Based Early Warning System for Dengue Fever}

\author{\IEEEauthorblockN{Julio Albinati}
\IEEEauthorblockA{Department of Computer Science\\
Universidade Federal de Minas Gerais\\
Belo Horizonte, Brazil 31270--901\\
Email: jalbinati@dcc.ufmg.br}
\and
\IEEEauthorblockN{Wagner Meira Jr}
\IEEEauthorblockA{Department of Computer Science\\
Universidade Federal de Minas Gerais\\
Belo Horizonte, Brazil 31270--901\\
Email: meira@dcc.ufmg.br}
\and
\IEEEauthorblockN{Gisele Lobo Pappa}
\IEEEauthorblockA{Department of Computer Science\\
Universidade Federal de Minas Gerais\\
Belo Horizonte, Brazil 31270--901\\
Email: glpappa@dcc.ufmg.br}}


%


\maketitle

\begin{abstract}
Dengue fever is a mosquito-borne disease present in all Brazilian territory. Brazilian government, however, lacks an accurate early warning system to quickly predict future dengue outbreaks. Such system would help health authorities to plan their actions and to reduce the impact of the disease in the country. However, most attempts to model dengue fever use parametric models which enforce a specific expected behaviour and fail to capture the inherent complexity of dengue dynamics. Therefore, we propose a new Bayesian non-parametric model based on Gaussian processes to design an accurate and flexible model that outperforms previous/standard techniques and can be incorporated into an early warning system, specially at cities from Southeast and Center-West regions. The model also helps understanding dengue dynamics in Brazil through the analysis of the covariance functions generated.
\end{abstract}


%
\IEEEpeerreviewmaketitle

\input{introduction}

\input{background}

\input{data}

\input{models}

\input{experiments}

\input{conclusions}

\section*{Acknowledgment}

This work was partially funded by projects InWeb (grant MCT/CNPq 573871/2008- 6), MASWeb (grant FAPEMIG/PRONEX APQ-01400-14), and by the authors’ individual grants from CAPES, CNPq and FAPEMIG.



\bibliographystyle{IEEEtran}
\bibliography{paper}
%
%
%

\end{document}

%% file: introduction.tex
\section{Introduction}

Dengue fever is a vector-borne disease transmitted by \emph{Aedes aegypti} \cite{world2009dengue}, the same mosquito that carries the viruses of zika, yellow fever and chikungunya.
According to \cite{samir2013global}, dengue fever is ubiquitous throughout the tropics, with highest risk zones in the Americas and Asia. This same study estimates that about 390 million people worldwide contract dengue fever every year, with 96 million of these cases being symptomatic. Although case fatality rate can be as low as 1\% with proper treatment, the disease causes great economical and social burden.

Dengue fever is endemic in most Brazilian regions, with its spread facilitated by suitable climatic and urbanization conditions. Brazil is the country with the largest number of dengue fever cases in the Americas, accounting for at least one-fourth of the symptomatic cases in the continent \cite{samir2013global}. Although the country has a successful surveillance program, it still needs an early warning system (EWS) capable of predicting outbreaks with high accuracy, so that health authorities may act in advance to minimize the impacts of the disease.

In this direction, previous attempts to model dengue fever incidence have exploited the existent relationship between climate and sociodemographic factors with dengue (\cite{naish2014climate,louis2014modeling}). These works usually employ relatively simple (parametric) models, such as (generalized) linear models (\cite{hu2012spatial, porcasi2012operative, chen2012modeling, earnest2012meteorological}) or (seasonal) autoregressive models (\cite{gharbi2011time, hu2010dengue}). Although these models provide explicit information about the relationship between dengue fever and covariates, they may fail to capture the complexity inherent to dengue dynamics. Besides that, most models cannot be used in EWS, as their main purpose is to \emph{explain} previous data, and not to \emph{predict} future dengue outbreaks. In fact, the survey in \cite{louis2014modeling} indicates that, out of 26 works, only 3 were classified as EWS.

In order to handle these issues, this paper proposes to model dengue fever incidence rate in Brazilian cities with Gaussian processes (GPs), a Bayesian non-parametric modelling framework that allows for flexible, accurate and interpretable models. Our main contribution is a model based on GPs equipped with a quasi-periodic kernel for predicting dengue fever incidence at city level with four weeks in advance. This time lag allows health authorities time to act to reduce the impact of future outbreaks.

The resulting model proved to be accurate and flexible, obtaining better results than previous approaches in the vast majority of the cities under study. The model is also interpretable and can provide interesting information about dengue dynamics in Brazil. Finally, although we focus on dengue fever in Brazil, we believe the model to be very general and applicable to distinct localities and diseases that exhibit seasonality.
 
This paper is organized as follows. Section \ref{sec:background} gives a brief background on GP regression. Next, we present our data collection. Section \ref{sec:model} introduces the proposed model, while Section \ref{sec:experiments} discusses  experimental results. Finally, Section \ref{sec:conclusion} draws our conclusions and indicates directions of future work.

%% file: background.tex
\section{Background} \label{sec:background}

Intuitively, a Gaussian process (GP) can be seen as an extension of the multivariate Gaussian distribution to the space of functions. GP is a distribution over functions parameterized by a a \emph{mean function} $\mu(\boldsymbol{x})$ and a \emph{covariance function} or \emph{(positive semidefinite) kernel} $k(\boldsymbol{x_i}, \boldsymbol{x_j})$. Placing a GP prior over a function implies that, for any finite set of points from the function's domain, their corresponding image will be jointly Gaussian-distributed.
 
GPs can be used to model many different learning tasks. Here we are interested in regression, where they are used as follows. We first assume that observations are noise-corrupted evaluations of a latent function $f(\boldsymbol{x})$. That is,
\begin{equation}
  y(\boldsymbol{x}) = f(\boldsymbol{x}) + \epsilon \mbox{, where } \epsilon \sim \mathcal{N}(0, \sigma^2)
  \label{eq:model}
\end{equation}

We then assume a GP prior over the latent function:
\begin{equation}
  f \sim \mathcal{GP}(\mu(\boldsymbol{x}), k_f(\boldsymbol{x_i}, \boldsymbol{x_j}))
  \label{eq:latent}
\end{equation}

Using \ref{eq:model} and \ref{eq:latent}, we conclude that $y(\boldsymbol{x})$ also follows a GP prior:
\begin{equation}
  y \sim \mathcal{GP}\left(\mu(\boldsymbol{x}), k(\boldsymbol{x_i}, \boldsymbol{x_j}) = k_f(\boldsymbol{x_i}, \boldsymbol{x_j}) + \sigma^2 \delta_{ij}\right)
\end{equation}
where $\delta_{ij}$ is the Kronecker delta, which is equals to 1 if $i = j$ and 0 otherwise.

Finally, given $N$ pairs of observations $(\boldsymbol{x}_i, y_i)$ and an unobserved point of interest $\boldsymbol{x}_*$, we have that
\begin{eqnarray}
  y_*|\boldsymbol{x}_*, X, \boldsymbol{y} & \sim & \mathcal{N}(\hat{\mu}, \hat{k}) \nonumber \\
  \hat{\mu} & = & \boldsymbol{k}_*^T (K + \sigma^2 I)^{-1} \boldsymbol{y} \nonumber \\
  \hat{k} & = &  k(\boldsymbol{x}_*, \boldsymbol{x}_*) - \boldsymbol{k}_*^T (K + \sigma^2 I)^{-1} \boldsymbol{k}_*
  \label{eq:pred}
\end{eqnarray}
where $X$ is an $N$-by-$D$ matrix composed of observed points, $\boldsymbol{y}$ is an $N$-length vector of observed targets, $\boldsymbol{k}_*$ is an $N$-length vector with the covariance between $\boldsymbol{x}_*$ and observed points, and $K$ is an $N$-by-$N$ matrix with the covariance between observed points.

By this definition, GPs allow us to obtain the exact predictive distribution through a closed-form expression. They are also flexible, since one can use any positive semidefinite kernel as a covariance function. Finally, they generate interpretable models, as one can interpret the covariance function as a measure of similarity between points, providing rich insights about the dependencies between them. For a more complete description, see \cite{Rasmussen:2005:GPM:1162254}.

%% file: data.tex
\section{Data Collection} \label{sec:data}

In this work, we use data provided by the Brazilian Ministry of Health regarding the number of confirmed dengue fever cases for the 298 Brazilian cities that have a population over 100,000 inhabitants. For each city, we have the weekly number of confirmed cases from January 2011 to December 2014, in a total of 209 weeks. In order to make results from cities with distinct population sizes comparable, we calculate the \emph{dengue incidence rate} (DIR) for each city, which is the number of cases of dengue fever per 100,000 inhabitants.

We also aggregate three climate covariates to our dataset: weekly average temperature, rainfall and humidity for the period under study. Data was collected from 287 meteorological stations associated with INMET\footnote{Brazilian National Institute of Meteorology - http://www.inmet.gov.br}.  Since meteorological stations are not present in every city, we associated each city to its closest geographical station.

In order to remove possible recording errors from the time series of confirmed dengue cases, we applied a pre-processing step to remove additive outliers (i.e., outliers of a time series that affect a single point in time) using the method described in \cite{chen1993joint}, which simultaneously estimates an autoregressive model and identifies outliers instead of using a fixed model fitted with all data.

\begin{figure}
  \centering
  \includegraphics[width=0.64\linewidth]{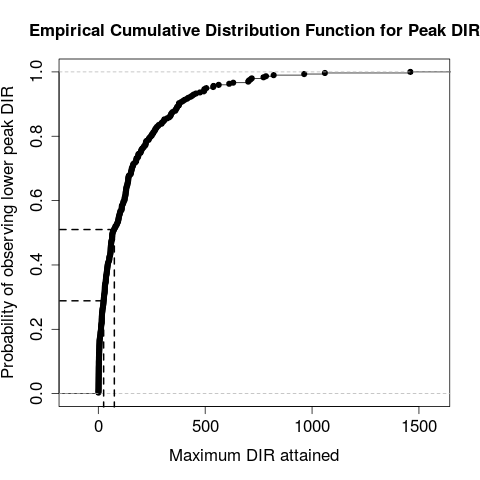}
  \caption{Empirical cumulative distribution for maximum DIR attained for all 298 Brazilian cities under study. Dashed lines indicate boundaries between incidence bands.}
  \label{fig:incidence}
\end{figure}

Figure \ref{fig:incidence} shows the cumulative distribution function for peak DIR, that is, the highest attained value of DIR for each city from 2011 to 2014. The Brazilian Ministry of Health defines three monthly incidence bands. Low incidence is defined as DIR less than 100, while medium incidence is defined for DIR between 100 and 300 and high incidence is defined as DIR greater than 300 in a month. Since we are dealing with weekly data, we divided these boundaries by four, arriving at 25 and 75. The dashed lines in Figure \ref{fig:incidence} show that around 49\% of the cities reached high incidence during the period under study, and 71\% reached at least medium incidence.

%% file: models.tex
\section{Proposed Model} \label{sec:model}

This section discusses the proposed non-parametric model for dengue incidence modelling. An initial issue in modelling count data -- such as the number of dengue cases in a city -- as a GP is that this kind of data is usually not appropriately modelled by a Gaussian distribution. Typically, we have two possible strategies to deal with this problem. The first strategy is to change the likelihood model, for instance by modelling the (logarithm transformed) mean function of a Poisson process or a negative binomial model as a Gaussian process. One example of this strategy can be seen in \cite{vanhatalo2007sparse}, where the authors used a Poisson likelihood model. The downside of this approach is that it loses some interesting properties of GPs, namely the exact inference and the closed-form calculations. Inference under this kind of model typically requires approximations, which may be computationally expensive. The second strategy is to apply an appropriate transformation on the response variable and model such transformation as a GP. In this work, we followed the second approach and used a logarithmic transformation on the response variable, which also increased the autocorrelation since it smooths very large outbreaks.

The main task in modelling via GPs is to define an appropriate covariance structure ($k_f$ in Equation \ref{eq:latent}), as it is standard practice in the literature to assume a zero-mean GP after centering the response variable. The covariance structure imposed by the GP prior should reflect what we expect from data. In the case of dengue fever incidence, we expect a smooth, possibly seasonal behaviour. We also expect to see a linear relationship between climate covariates and dengue incidence. Therefore, we define a three-part kernel to capture each of these assumptions. Let $\boldsymbol{x}_i$ denote a vector containing all climate covariates associated to week $i$ and $\boldsymbol{x}_j$ denote a similar vector associated to week $j$. Then, the covariance between $\boldsymbol{x}_i$ and $\boldsymbol{x}_j$ is given by
\begin{eqnarray}
  k_f\left(\boldsymbol{x}_i, \boldsymbol{x}_j\right) & = & k_{1}(\Delta t) + k_{2.1}(\Delta t) k_{2.2}(\Delta t) + k_{3}(\boldsymbol{x}_i, \boldsymbol{x}_j) \label{eq:cov} \\
  k_{1}(\Delta t) & = & \sigma_{loc}^2 \left( 1 + \frac{\sqrt{5}\Delta t}{\ell_{loc}} + \frac{5\Delta t^2}{3 \ell_{loc}^2} \right) exp \left( - \frac{\sqrt{5}\Delta t}{\ell_{loc}} \right) \nonumber \\
  k_{2.1}(\Delta t) & = & \sigma_{qp}^2 \left( 1 + \frac{\sqrt{5}\Delta t}{\ell_{qp}} + \frac{5\Delta t^2}{3 \ell_{qp}^2} \right) exp \left( - \frac{\sqrt{5}\Delta t}{\ell_{qp}} \right) \nonumber \\
  k_{2.2}(\Delta t) & = & exp\left(-2 sin^2\left( \frac{\pi \Delta t}{p} \right) / \ell_{per}^2 \right) \nonumber \\
  k_{3}\left(\boldsymbol{x}_i, \boldsymbol{x}_j\right) & = & \sigma_{lin}^2 + \boldsymbol{x}_i^T Diag(\ell_{rain}^2, \ell_{temp}^2, \ell_{hum}^2)^{-1} \boldsymbol{x}_j \nonumber
\end{eqnarray}
where $\Delta t = |i - j|$ is the absolute distance in weeks between points $\boldsymbol{x}_i$ and $\boldsymbol{x}_j$, $Diag$ is a diagonal matrix formed of its arguments, $\sigma_{loc}$, $\sigma_{qp}$, $\ell_{loc}$, $\ell_{qp}$, $\ell_{per}$, $\ell_{rain}$, $\ell_{temp}$, $\ell_{hum}$ and $p$ are hyperparameters learned from data via likelihood (local) maximization.

The first component of the kernel ($k_{1}(\Delta t)$) is a Mat\'ern kernel over the time dimension, used to enforce the assumption that following weeks should be correlated. Its hyperparameter, $\ell_{loc}$ and $\sigma_{loc}$, are used to control the range of weeks that should correlate and the strength of the correlation signal, respectively. The second component is a multiplication of a Mat\'ern kernel ($k_{2.1}(\Delta t)$) and a periodic kernel ($k_{2.2}(\Delta t)$). It is used to exploit the yearly periodicity observed in dengue incidence in some cities, while still being able to give more importance to closer years. The hyperparameters $\ell_{qp}$, $\ell_{per}$, $p$ and $\sigma_{qp}$ are useful for controlling the decay of quasi-periodicity (how many past periods should impact the incidence in a given week), the decay of periodicity (how fast the periodic signal should decay), the period and the strength of this correlation signal, respectively. Finally, the third component ($ k_{3}(\boldsymbol{x}_i, \boldsymbol{x}_j)$) is a linear kernel over weather-related covariates, where the diagonal matrix $M$ is used to control the impact of each covariate and $\sigma_{lin}$ allows for a bias term.

%% file: experiments.tex
\section{Experimental Evaluation} \label{sec:experiments}

\begin{figure*}[t!]
  \centering
  \includegraphics[width=0.75\linewidth]{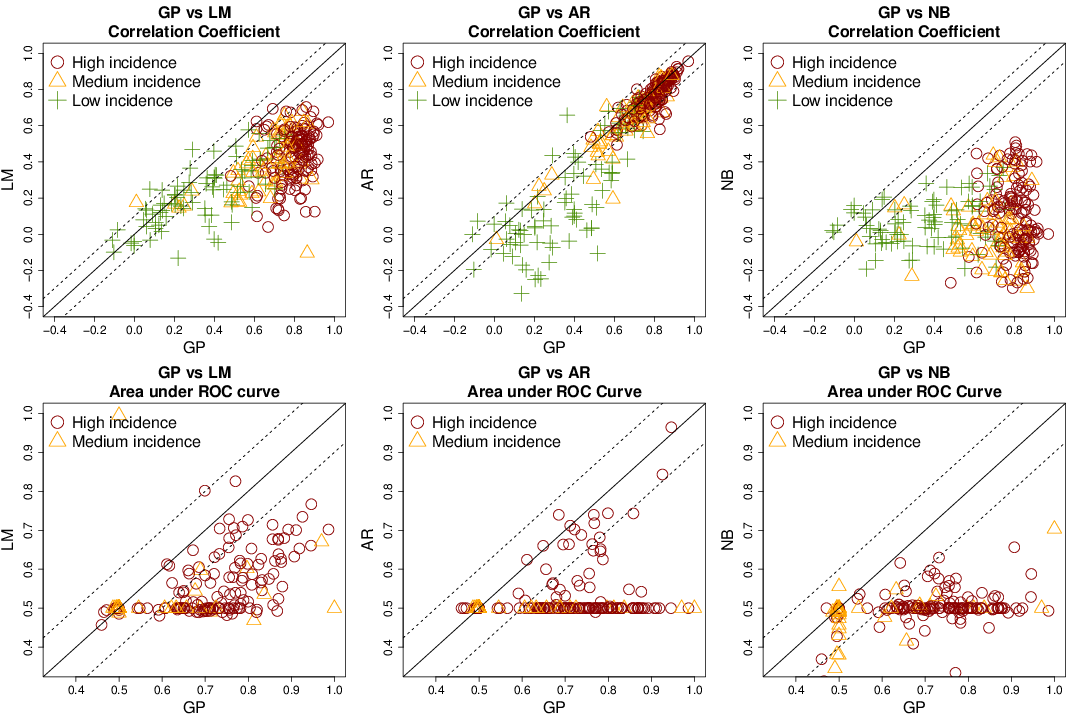}
  \caption{Comparison between GP and baselines according to both accuracy metrics. Each point represents a city, where form and color indicates the highest incidence band reached by such city. Points below the solid diagonal line indicates that GP performed better, while points above the solid diagonal line indicates better performance of the baseline. Points below or above the dashed diagonal line indicates difference in performance greater than 0.1.}
  \label{fig:baselines}
\end{figure*}

This section evaluates the proposed model in 298 Brazilian cities. Predictions were always made with 4 weeks in advance, starting at the beginning of the third year (week 105). Weeks up to the end of the second year were used only as training. Therefore, in order to predict week $t$, we used as training data from week 1 to $t-4$, for $t = 105, 106, ..., 209$.

Climate covariates were normalized to have zero mean and unit variance.
Such covariates are usually used with some lag in the dengue modelling literature (\cite{naish2014climate,louis2014modeling}), as we expect it to interact with the life cycle of dengue's vector. Thus, after some excessively hot and/or humid period, we expect to observe a proliferation of mosquitoes that would then be able to acquire the virus and infect humans. This process requires some time, so climate covariates and dengue fever time series are better aligned if we shift the former some weeks into the future. For each city and each covariate, we calculated the optimal lag by computing the correlation between a covariate and target values when applying lags from 4 to 26 weeks, using only training data (the first and second years).

Models were evaluated using two performance metrics: (i) Pearson correlation coefficient between predicted and real values and (ii) area under the ROC curve (AUC) using the weekly incidence bands defined in Section \ref{sec:data}. 
The former was used instead of error-based metrics because it is scale-independent and allows for easier comparison.
For the latter, we only use cities that reached medium and/or high incidence, since 
all low incidence cities only have low incidence throughout the whole period. 
Comparison between models was done using Wilcoxon test with a confidence level of 95\%, using Bonferroni correction if applicable.

\subsection{Comparison with Baselines}

In order to assess the accuracy of the proposed model, we compare it with three baselines: a linear model, an autoregressive model and a negative binomial model specifically proposed to model dengue fever in Brazil (\cite{lowe2010spatio}).

The linear model (LM) uses the same climate covariates indicated previously. The response variable is also previously transformed via a logarithmic link function.

The autoregressive model (AR) is of first order, making predictions for a week $t$ using information from week $t-1$. We obtained predictions within 4 weeks in advance by applying successive predictions. A window of 12 weeks was used, so that we train the model for each week ignoring data from more than 3 months into the past. Again, a logarithmic transformation was previously applied.

The negative binomial (NB) proposed in \cite{lowe2010spatio} is a generalized linear mixed model. It was originally proposed to deal with monthly data at microregion (a group of geographically close municipalities) level. Since we are dealing with weekly data at city level, we adapted it by using weekly covariates and defining neighbouring cities as cities within 500 km of each other. The model uses as predictors temperature and precipitation (both as a 13-week average, lagged 8 weeks), Oceanic Ni\~{n}o Index (ONI, lagged 26 weeks), population density and altitude. The random effects include an unstructured spatial effect to allow spatial heterogeneity, a conditional autoregressive (CAR) effect to allow dependences between neighbouring cities and a temporal autoregressive effect. Letting $\mu_t^{(i)}$ and $\kappa$ denote the mean DIR value for city $i$ during week $t$ and the scale parameter, respectively, the model is given by
\begin{eqnarray}
  y_t^{(i)} & \sim & NegativeBinomial\left(\mu_t^{(i)}, \kappa\right) \label{eq:rachel} \\
  log\left(\mu_t^{(i)}\right) & = & log\left(e^{(i)}\right) + \alpha + \sum_j \beta_j x_{jt}^{(i)} + \phi^{(i)} + \nonumber\\ & & \nu^{(i)} + \omega_{Month(t)} \nonumber
\end{eqnarray}
where $y_t^{(i)}$ indicates the DIR during week $t$ at city $i$, $e^{(i)}$ is the expected dengue risk at city $i$ and is based only on the city's population size, $\boldsymbol{x}^{(i)}_t$ is a vector formed of all covariates associated to week $t$ and city $i$ and $x_{jt}^{(i)}$ indicates the $j$-th position of $\boldsymbol{x}^{(i)}_t$.
For the formal definition of the random variables in Equation \ref{eq:rachel}, see \cite{lowe2010spatio}.

Inferences under this model, however, require expensive Markov Chain Monte Carlo sampling techniques if all cities under study were used. In order to reduce computational effort, we obtained 35 clusters of up to 10 cities and estimated the parameters for each cluster. Previous experiments indicated that clustering cities based on the correlation between each pair of cities on a hierarchical fashion was the best strategy. However, we stress that allowing larger clusters did not lead to a statistically significant improvement of the results obtained.

Figure \ref{fig:baselines} shows the comparison between the proposed model (GP) and baselines. Observe that GP obtained better results than all baselines when considering both accuracy metrics, outperforming them in at least 78\% of the cities when considering the correlation coefficient or AUC. 
A Wilcoxon test indicated that GP obtained more accurate predictions than all baselines with a confidence level of 95\%. The correlation coefficient indicates that GP was better than LM on medium/high DIR cities, with similar results on low DIR cities, while the opposite happened when compared to AR. When considering the AUC, cities where GP is outperformed by some baseline are cities where both models do not work well (AUC close to 0.5). For those cities, the typical DIR during dengue season stays close to the boundaries, thus being harder to predict.

\subsection{Analysis of Hyperparameters}

Interpretability of GPs come from the hyperparameters of the covariance function, which tell us how data points are related. In this section, we study the hyperparameters obtained after optimization, shown in Table \ref{tab:hyper}. In order to make interpretation easier, some values were reported squared.

\begin{table}[h!]
  \centering
  \caption{Hyperparameters Obtained After Likelihood Maximization}
  \label{tab:hyper}
  \begin{tabular}{cccc}
    \hline
    & 1st Quartile & Median & 3rd Quartile \\
    \hline
    $\sigma_{loc}^2$ & 0.039 & 0.101 & 0.186 \\
    $\sigma_{qp}^2$ & 0.484 & 1.427 & 2.311 \\
    $\sigma_{lin}^2$ & 0.005 & 0.022 & 0.082 \\
    $\ell_{loc}$ & 1.000 & 2.000 & 7.000 \\
    $\ell_{qp}$ & 41.000 & 58.000 & 99.000 \\
    $\ell_{per}$ & 1.000 & 1.000 & 2.000 \\
    $\ell_{rain}^2$ & 558.000 & 2675.000 & 10270.000 \\
    $\ell_{temp}^2$ & 164.000 & 768.000 & 4316.000 \\
    $\ell_{hum}^2$ & 493.000 & 2449.000 & 9879.000 \\
    $p$ & 54.000 & 59.000 & 73.000 \\
    \hline
  \end{tabular}
\end{table}

The hyperparameters $\sigma_{loc}$, $\sigma_{qp}$, $\sigma_{lin}$ tell us the importance of each component of our kernel. The first and the second are the highest value that the local and quasi-periodic component can achieve, respectively, while the third is the value given by the linear component when climate covariates are exactly at their mean value. Notice that the quasi-periodic component has the highest impact in the kernel, confirming the notion that we should explore information from past years' data.

The hyperparameters $\ell_{loc}$ and $\ell_{qp}$ indicate how fast is the decay of the covariance signal of the local and quasi-periodic components, respectively. One way to interpret these values is to think that if the difference in time between a pair of points is smaller than $\ell_{*}$, then the covariance signal from a given component will be greater than $\sigma^2_{*}/2$, that is, greater than half of the component's maximum attainable value. For the local component, only points up to two weeks away from each other were highly correlated for most of the cities. For 25\% of the cities, however, more distant relationships were found to be relevant. On the other hard, the quasi-periodic component typically correlates points from subsequent years (52 weeks). For some cities, though, even points separated by two years were correlated. The period, given by $p$, also plays a major role on the quasi-periodic component. As expected, most of the cities exhibited yearly periodicity. For at least 25\% of the cities, the periodicity was found to be greater than 52, indicating some shift on the typical dengue fever season.

Finally, $\ell_{rain}$, $\ell_{temp}$, $\ell_{hum}$ provide the relative importance of each covariate. The higher the value, the less important a given covariate is. We see that rainfall was found to be the least important covariate, while temperature proved to be the most relevant one. However, since the values of all these hyperparameters are high, we observe that climate covariates do not add much information to the model. This is probably due to the fact that climate in Brazil also exhibits a yearly periodicity, thus being captured by the quasi-periodic component.

It is interesting to notice the variability shown in Table \ref{tab:hyper}, indicating the flexibility of the model. It can automatically shift from a highly periodic scenario to a scenario where local information is more important, thus accommodating a wide range of behaviours within a single model.

\subsection{Analysis of Predictions}

\begin{figure}[!h]
  \centering
  \includegraphics[width=0.8\linewidth]{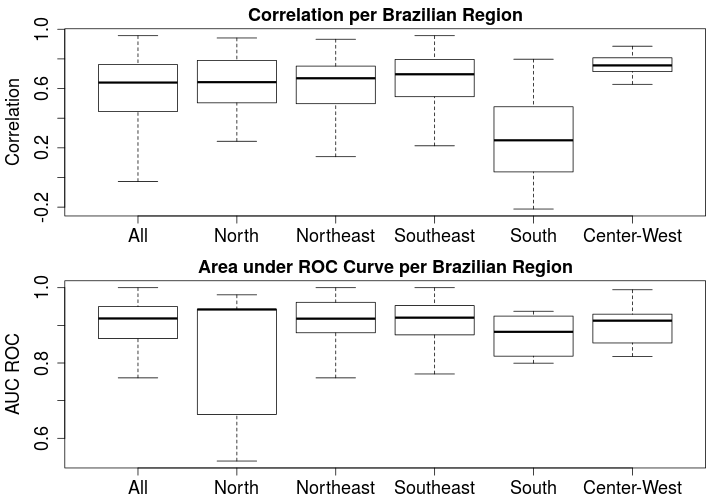}
  \caption{Results obtained when considering all cities and filtering by each of the five Brazilian regions.}
  \label{fig:regions}
\end{figure}

\begin{figure}[!h]
  \centering
  \includegraphics[width=0.64\linewidth]{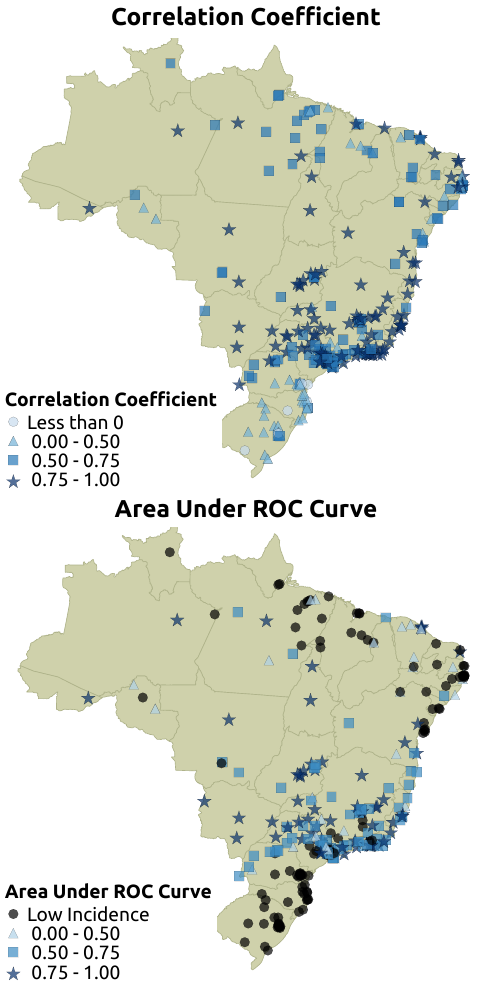}
  \caption{Spatial distribution of correlations and areas under ROC curve obtained. Each dot represent a city and its color and shape indicates its accuracy according to a metric. Black dots on the bottom map indicates cities that never reached medium and/or high incidence and were not evaluated according to AUC.}
  \label{fig:maps}
\end{figure}

In this section, we analyse the results obtained by the proposed model in more detail. Figure \ref{fig:regions} shows the results obtained according to the correlation coefficient and AUC for all cities under study, as well as results for each of the five Brazilian geographical regions. When considering the correlation coefficient, the Southeast and Center-West regions obtained the best results, with median correlation of 0.70 and 0.76, respectively, which are greater than the median correlation of all cities, 0.64. When considering AUC, all cities obtained similar results, with a global median of 0.92. This analysis is strengthened by Figure \ref{fig:maps}, which shows the spatial distribution of results obtained by both metrics. Note that most cities from North, Northeast and South exhibit low incidence of dengue fever (black dots on bottom map of Figure \ref{fig:maps}), while cities from Southeast and Center-West typically have higher DIR. This leads us to conclude that the proposed model obtains better results on cities with higher DIR. In fact, data from low incidence cities are usually much noisier, thus being naturally harder to predict.

\begin{figure*}[!t]
  \centering
  \includegraphics[width=0.91\linewidth]{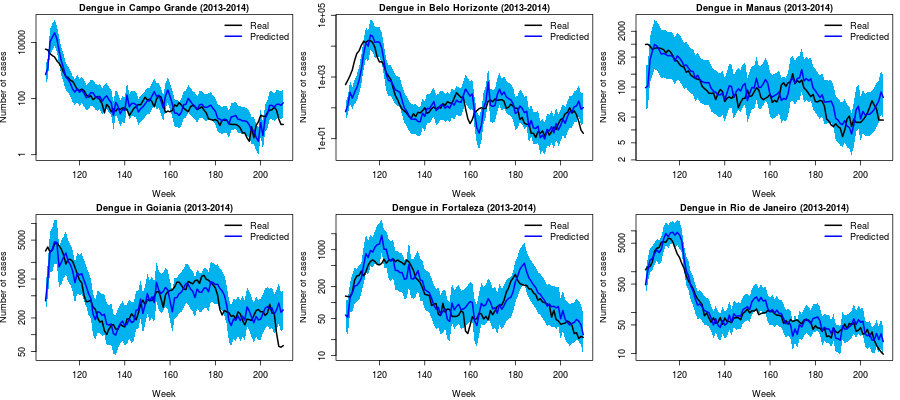}
  \caption{Predictions for the six Brazilian capital cities with highest DIR. The black line indicates the real number of confirmed cases, while the blue line indicates the prediction given by the proposed model. The blue shaded area shows the 95\% confidence interval.}
  \label{fig:examples}
\end{figure*}

Finally, Figure \ref{fig:examples} provides a more qualitative analysis of the model, showing predictions for the six Brazilian capital cities with highest DIR. Predictions follow a similar pattern to real values, which are almost within the confidence interval.

%% file: conclusions.tex
\section{Conclusions and Future Work} \label{sec:conclusion}

In this work, we propose to model DIR at Brazilian cities as a GP equipped with a quasi-periodic covariance function. This function allows the model to exploit both local and distant information in order to accurately predict DIR.

We showed that the proposed model outperforms a previous model specifically design to dengue fever in Brazil. This gain in accuracy is due to the non-parametric nature of the proposed model and its flexibility, thus being able to automatically adapt to different scenarios. This reasoning is strengthened by the analysis of hyperparameters, which differs from one city to another to capture distinct patterns.

The proposed model obtained the best results on cities from Southeast and Center-West regions of Brazil, which exhibit the highest DIR. Therefore, the model can be used as an EWS, since it performs well where it is mostly needed.

We would like to highlight that we believe the proposed model to be quite general and applicable to other diseases that exhibit seasonality, which includes virtually all mosquito-borne diseases. In this sense, it is an initial step towards defining accurate EWS for a large number of diseases. As future work, we intend to extend the model to incorporate spatial dependencies. It is known that dengue outbreaks are not geographically isolated events, so that an outbreak in a city may increase the possibility of outbreaks at other cities.